\documentclass[fleqn,12pt,twoside]{article}
\usepackage[headings]{espcrc1}
\usepackage{graphicx}

\def\combin#1#2{\ensuremath{{\left(\begin{array}{c} #1 \\ #2
\end{array}\right)}}}
\def\sigmain{\ensuremath{{\sigma_{\rm{in}}}}}
\def\av#1{\ensuremath{\langle#1\rangle}}
\def\th{\ensuremath{^{\rm{th}}}}

\title{Quantum Optics and Heavy Ion Physics}

\author{Roy J. Glauber\\ \medskip
Lyman Laboratory of Physics, Harvard University \\
Cambridge, MA 02138, USA }

\runauthor{R. Glauber}

\begin{document}

\maketitle

\begin{abstract}
I shall try to say a few words about two particular ways in which
my own work has a certain  relation to your work with heavy ions.
My title is therefore \emph{quantum optics and heavy ion physics}.
\end{abstract}

Some of this work deals with a form of nuclear scattering theory.
The approach I shall talk about is really a \emph{nuclear}
diffraction theory~\cite{1}.  It grows directly out of the ancient
Fraunhofer form of optical diffraction theory, which in effect is
a theory of small momentum transfer collisions. The ``collisions"
take place between a plane-wave and a screen of some sort with an
aperture in it, or with a partially absorbing obstacle. These
waves in effect describe small momentum transfer collisions
because the deflections of the wave are usually small. Most of the
scattered intensity is thereby cast near the forward direction.

\begin{figure}
\begin{tabular}{lccr}
\textit{a}:\includegraphics[width=160pt]{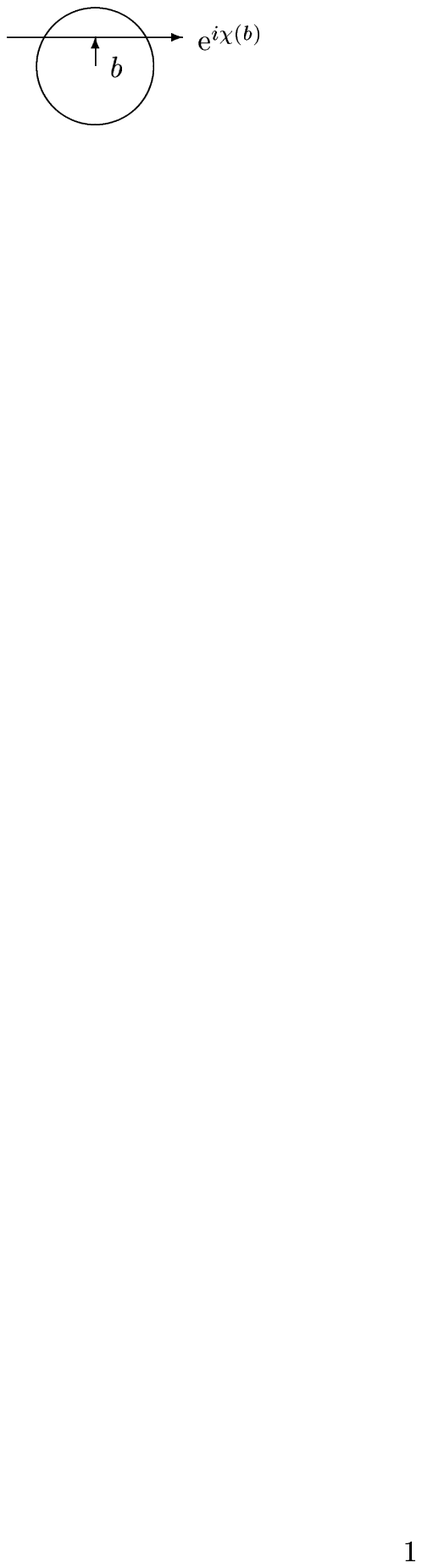}&&&
\textit{b}:\includegraphics[width=160pt]{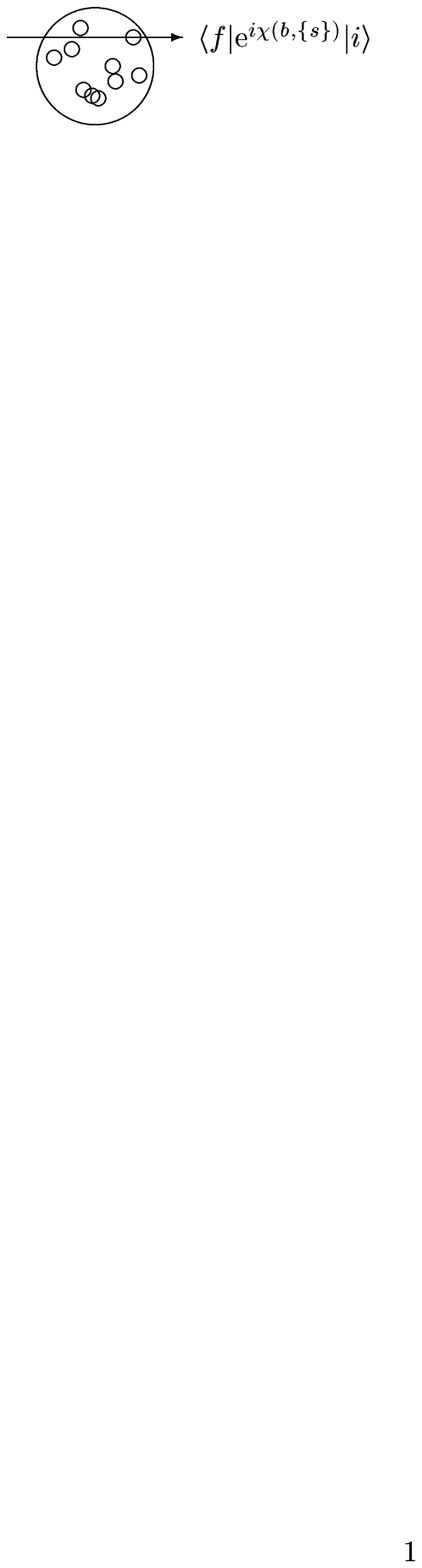}\\
\end{tabular}
  \caption{Illustration of the optical model. Fig~\ref{f:opt}\textit{a}: The wave travelling
through the nucleus suffers changes of phase more than amplitude.
Here the phase-shift is $\chi(b)$ at the impact parameter $b$.
Fig~\ref{f:opt}\textit{b}: The nucleus makes a transition from an
initial state $|i\rangle$ to a final state $|f\rangle$. The energy
change has to be small compared to the incident
energy.}\label{f:opt}
\end{figure}

There is of course always a requirement that the wave-length be
small compared to the dimensions, whatever they may be, of the
aperture or the obstacle. Now how does one apply this approach in
nuclear physics? The most elementary and passive model of the
nucleus is as a transparent sphere of some sort, perhaps a cloudy
or partially absorbing one. That is the so-called optical model.
The wave travelling through that sphere suffers changes both of
amplitude and phase. Here in Fig.~\ref{f:opt}\textit{a} we have
such a phase-shift $\chi(b)$ at the impact parameter $b$. Now
since deflections of the wave within the interaction region are
negligible, all of the scattered intensity is thrust  near the
forward direction, and that happens to be where the approximation
is a good one. In fact it is for that reason a unitary
approximation, and it therefore possesses a certain
self-consistency.  In the case of elastic scattering, for example,
you can calculate the same total cross-sections either by
integrating  the approximate intensity over angles, or by using
the ``optical theorem."

Now if we want to become a little bit more adventurous, we can
introduce the transverse coordinates of the individual nucleons
${\bf s}_j(j = 1\dots A)$ within the nucleus. Again, we must
assume that the wavelength is small compared with all of the force
ranges in question. But now we can let the nucleus be less
passive. We can let it make a transition from an initial state
$|i\rangle$ to a final state $|f\rangle$. (See
Fig.~\ref{f:opt}\textit{b}.) We have to require that the energy
change be small compared to the incident energy, and in that case,
what we do is to take a matrix element of the same phase-shift
exponential $\exp\left[i\chi(b,\{{\bf s}\})\right]$ taking careful
account of its dependence on the internal coordinates. If you do
this in particular when the final state of the nucleus is the same
as the initial state, then you are dealing with elastic
scattering, and in that way you can actually derive the
expressions for the optical model  of the nucleus.

\begin{figure}
\begin{tabular}{llcrr}
\textit{a:} & 
\includegraphics[width=170pt]{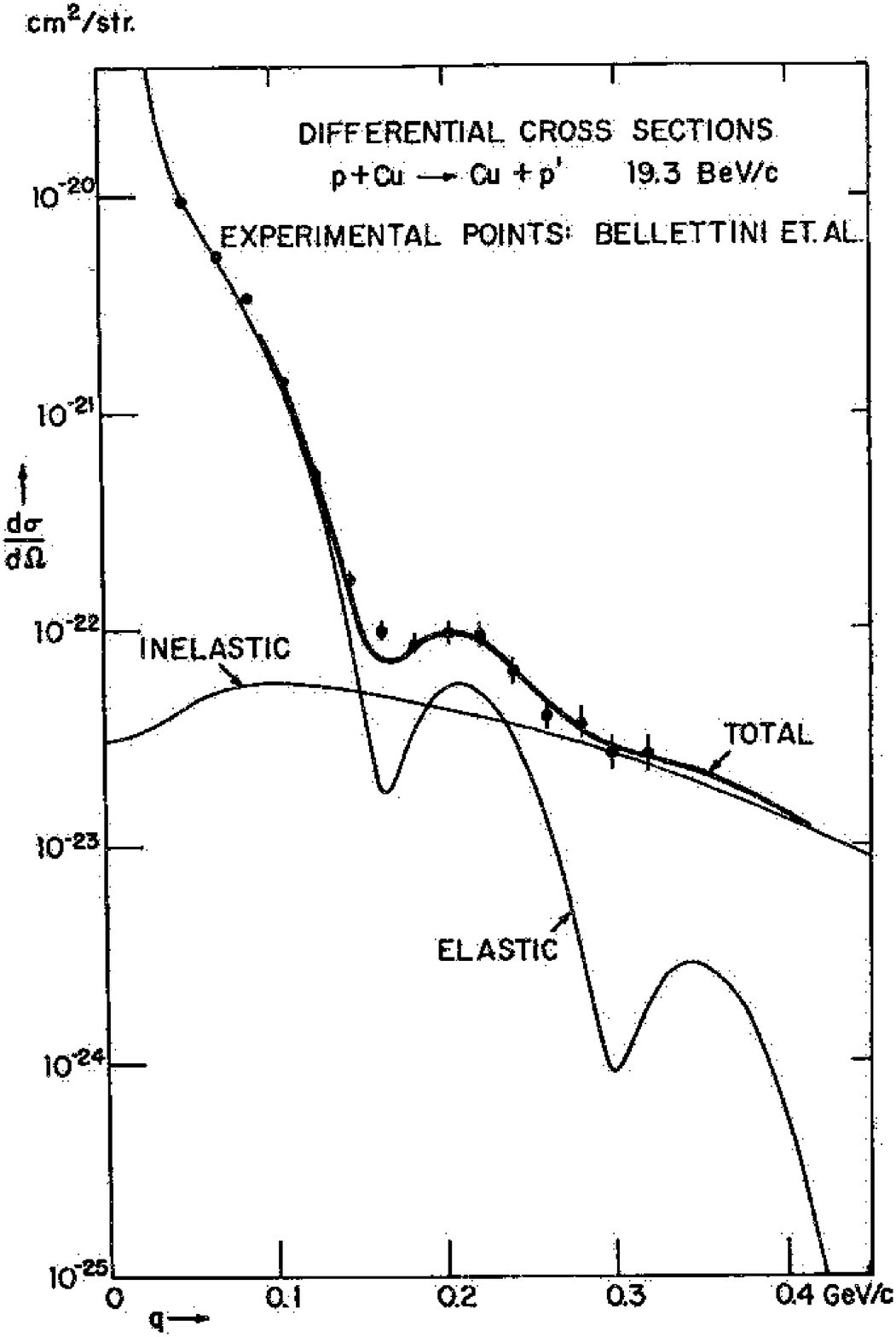} & 
$\qquad$ &
\textit{b:} &
\includegraphics[width=170pt]{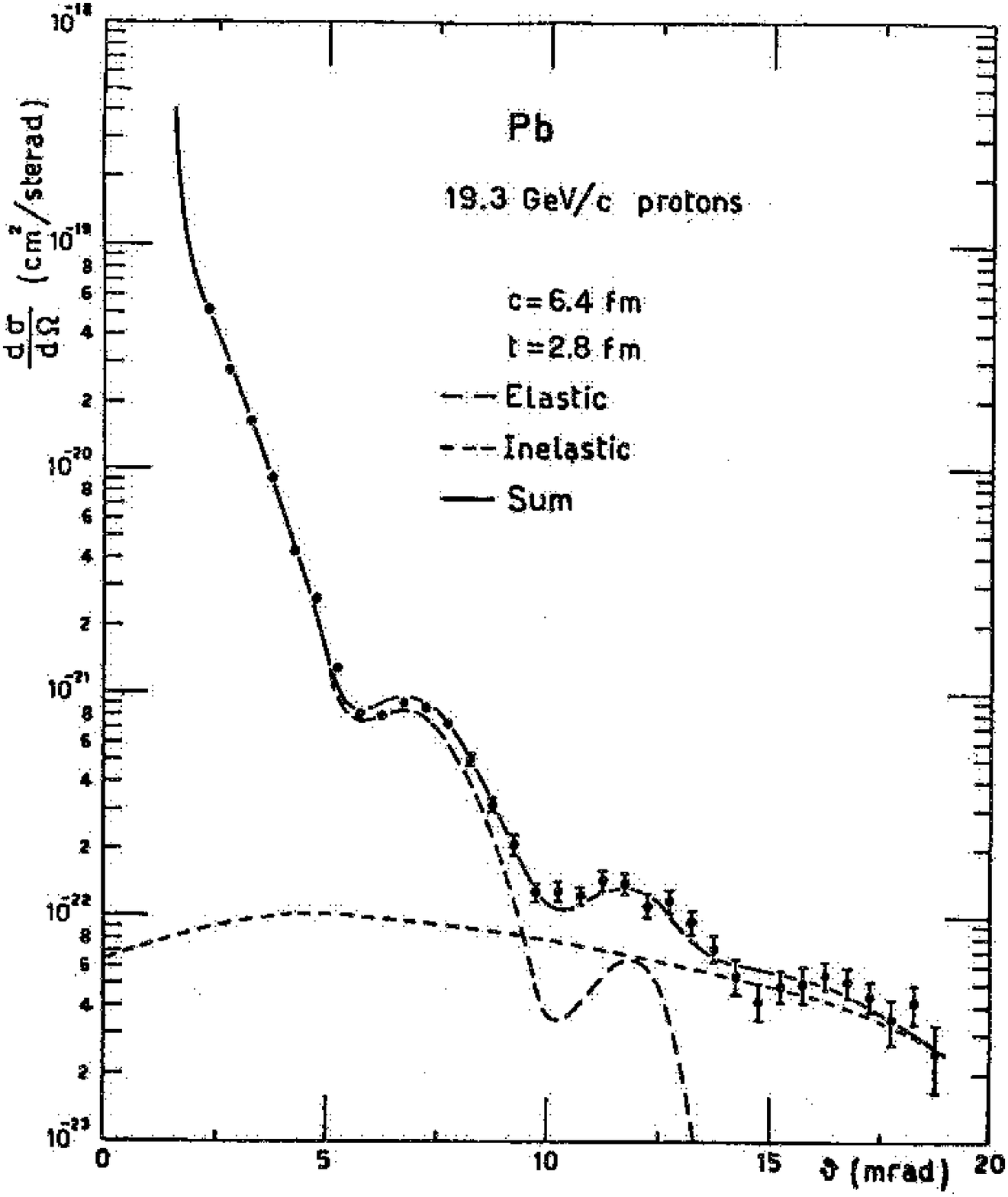}
\end{tabular}
  \caption{Elastic, inelastic and summed scattering
calculated by Kofoed-Hansen for 19.3 GeV/$c$ protons incident on
Cu~(\textit{a}) and by Matthiae for 19.3 GeV/$c$ protons incident
on Pb~(\textit{b}). Experimental points from Bellettini et
al~\cite{2}.}\label{f:elast}
\end{figure}

Now it may be, that the projectile particle has internal
coordinates too. A deuteron projectile, for example, consists of
two particles, and we can use the same scheme to treat transitions
that take place within the deuteron as well. That is one of the
things I want to tell you about. It is reasonable now to ask about
the accuracy of this simple theory. The diffraction theory was
found in the 1960's to work really quite well. You can see in
Fig.~\ref{f:elast}\textit{a} for 19.3 GeV/$c$ protons incident on
Cu a typical diffraction pattern for elastic scattering. It was
also possible to calculate the inelastic scattering and add that
to the elastic scattering to compare with the measured sum. The
experimental points show the results found in that era by the
Cocconi group at CERN. The theory seems to work well for all of
the elements. Shown in Fig.~\ref{f:elast}\textit{b} are the
measurements on lead. Again, the inelastic and the elastic
scattering are both represented well.

\begin{figure}
\textit{a}:\includegraphics[width=210pt]{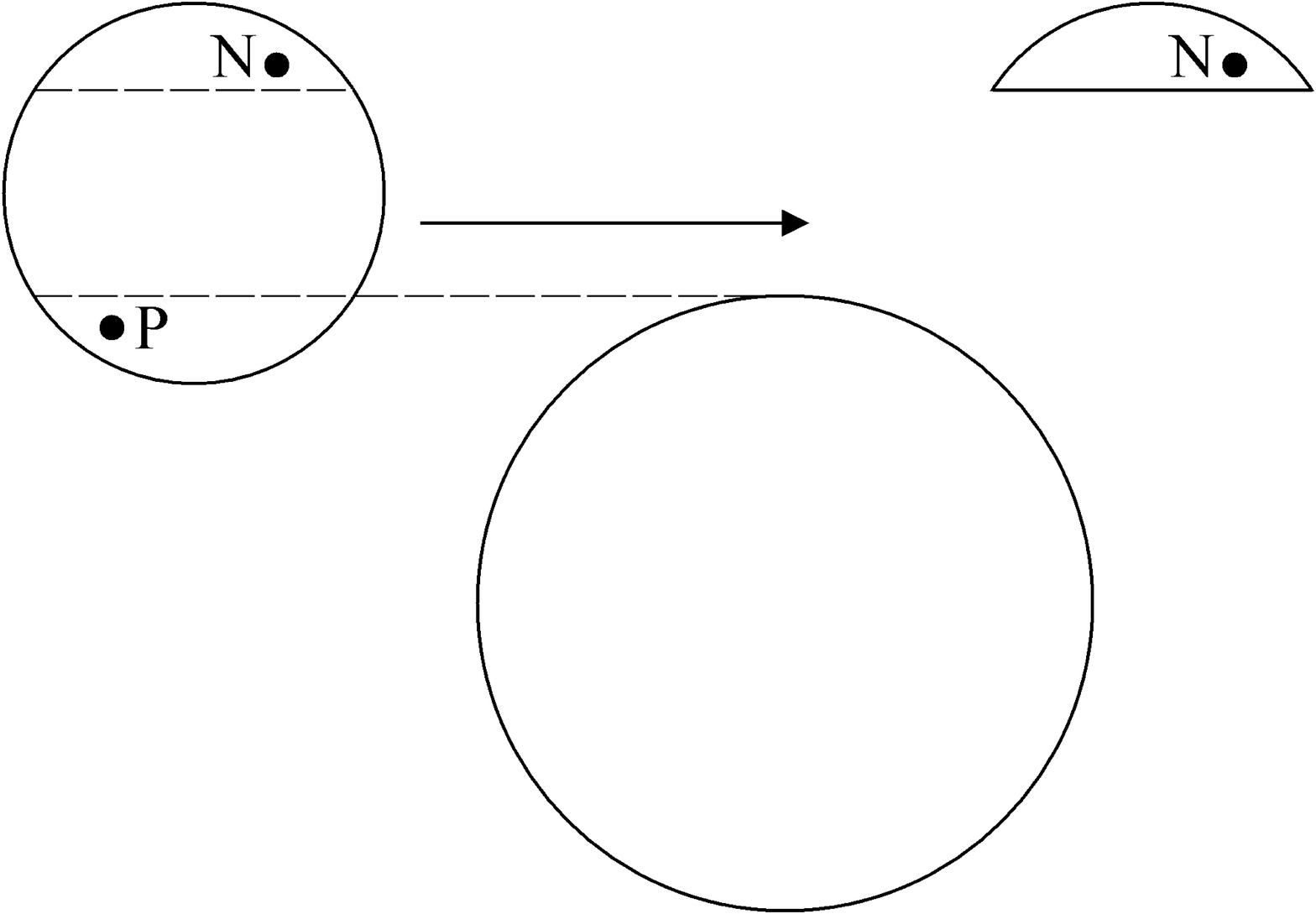}
\textit{b}:\includegraphics[width=210pt]{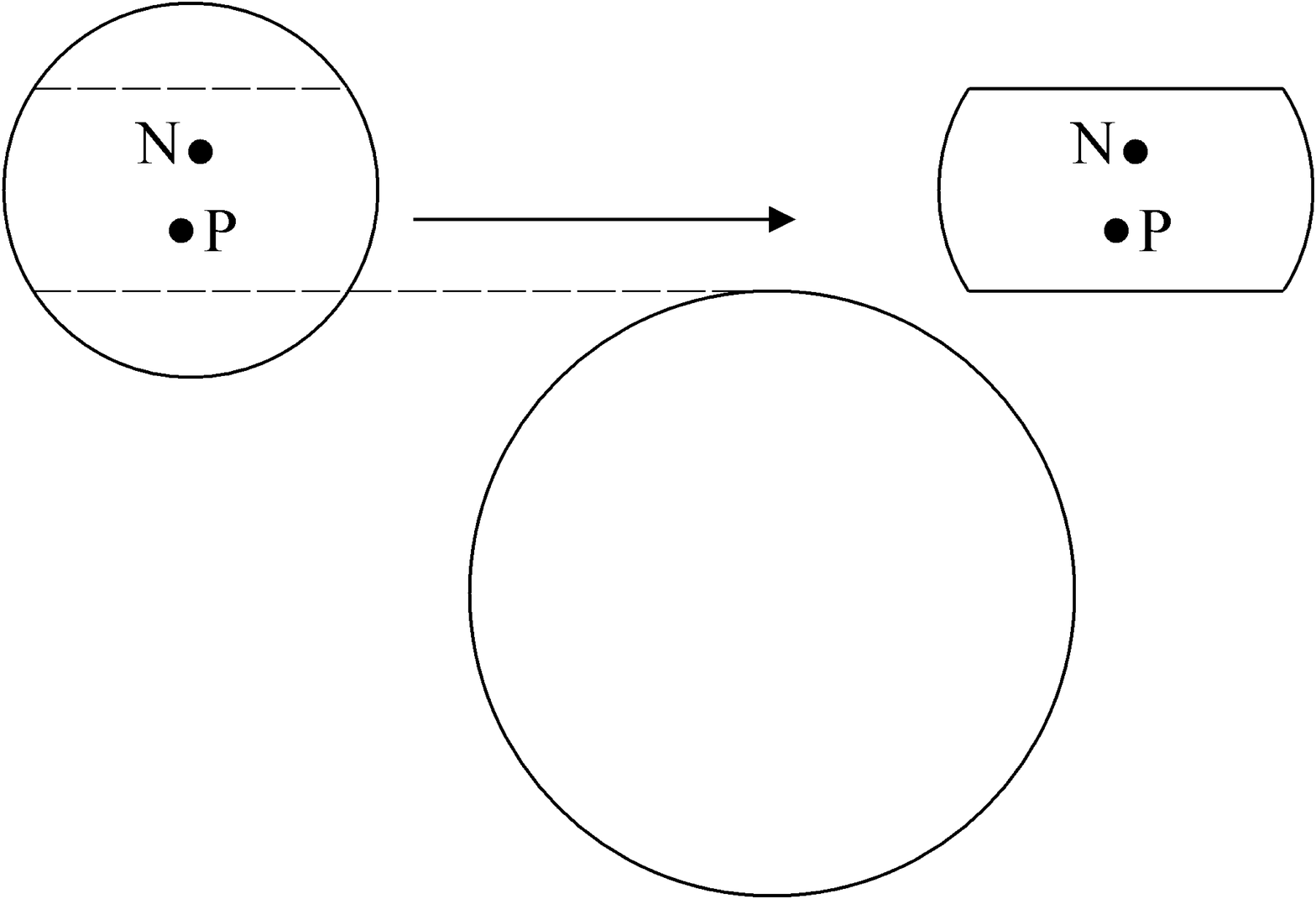}\\
\caption{Illustration of a deuteron-nucleus collision. The
deuteron is coming in from the left, grazing the edge of the
nucleus. When the neutron and proton happen to be far
apart~(\textit{a}), the proton strikes the nucleus and is stripped
away. The neutron partner will be left alone and keep moving on.
If the neutron and the proton are at the moment of encounter
closer together~(\textit{b}), then nothing hits the nucleus, both
the neutron and proton go forward, undeflected. A truncated
version of the deuteron wave-function is left.}\label{f:hic}
\end{figure}

I mention all of this because I want to tell you about another
kind of application which is somewhat less elementary.
Historically, however, it is a little bit older than the
scattering measurements. When they began operating the 184''
Berkeley cyclotron in 1947 they projected a beam of  deuterons
against  a target of heavy nuclei. The first thing observed was,
that  when they sent in a beam of 180 MeV deuterons, they got out
a forward travelling beam of 90 MeV neutrons. How did that happen?
Well, Robert Serber gave a very simple explanation~\cite{3}. Here
in Fig.~\ref{f:hic}\textit{a} is the deuteron coming in, and as
you see, it is just going to graze the edge of the nucleus. One
must always assume that the intranuclear motions are slow, as
indeed they are in the weakly bound deuteron. Now when the neutron
and proton happen to be far apart,the proton is going to strike
this opaque nucleus and be stripped away.  What then will happen?
Well, the neutron partner will be left alone and keep moving on.
This is the process that Serber called \emph{stripping}. (He
actually named it after something that had been done in Los Alamos
in a very different connection several years earlier.) There  was
something unknown at the time that Serber omitted from the
calculation, and that is what has since become known as
diffraction dissociation. Suppose the neutron and the proton
happen at that moment of encounter to be closer together, as in
Fig.~\ref{f:hic}\textit{b}. Well, then nothing hits the nucleus,
both the neutron and proton go forward, undeflected. But what is
left, populated still by both the neutron and the proton, is a
truncated version of the deuteron wave-function. Now the
neutron-proton system has only one bound state, the deuteron
ground state, so anything you do that chews away a part of the
deuteron wave-function is going to leave a certain component of
excited states present. As far as the excited state content of
this funny wave-function is concerned, the neutron and proton will
simply part company and separate. That describes the process of
diffraction dissociation, and it was first noticed just 50 years
ago~\cite{4}, so we have now something of an anniversary. I should
say that Good and Walker shortly after that, adopted this process
as a kind of model for the behavior of the nucleons themselves at
higher energies, and they gave it the name diffraction
dissociation~\cite{5}.  I too was thinking in those days about
diffractive models for what neutrons and protons do when they
produce small-angle sprays of particles in the forward direction.
But if you are a theorist and try to say anything about this, you
need a theory to back it up. Good and Walker were experimenters.
They did not need any theory, they simply assumed that there is a
process called diffraction dissociation, based on the analogy with
the deuteron.

Now I'd like to persuade you that there are some things that go on
in the diffraction theory that are not altogether obvious and have
some relevance to work on heavy ion collisions. I am sorry to say,
however, most of the things that are in this diffraction theory of
nuclear collisions, are omitted from your heavy ion work, because
you are ordinarily not very interested in small momentum transfer
processes.  Particle production processes, for example, typically
involve large momentum transfers.  So, after these omissions what
is left is simply the geometrical substrate of diffraction theory,
a kind of skeleton, and I find that my name is mentioned
frequently in connection with an extremely nasty multiple integral
that has to be done numerically. Well, this integral seems
principally to take two different forms, I find in attending some
of your lectures. One of these treats all collisions equally and
simply sums up the total number of collisions that takes place.
The other form was invented by some Polish friends of mine and I
will talk about that in a moment.

Let us first talk about  a single particle colliding inelastically
with a nucleus having A constituents. We want to derive the cross
section for  all possible combinations of individual inelastic
processes, whatever they may be:

\begin{equation}
\sigma_{\rm{part. prod.}} = \int \left\{1-(1-\sigmain
t(b))^A\right\}d^2b\,.
\end{equation}

Here \sigmain~is the relevant inelastic cross-section for
nucleon-nucleon collisions. The function $t$ at the impact
parameter $b$ is the effective thickness of a single nucleon and
it is normalized to unity,  $\int t(b)d^2 b = 1$. Now what is the
probability $P_n$ that $n$ of these inelastic collisions take
place? Well, you have to pick out a combination of $n$ nucleons
from the $A$ of them. The number of ways you can do that is given
by a familiar combinatorial coefficient, with the result that:
\begin{equation}
P_n = \combin{A}{n} (\sigmain t)^n (1-\sigmain
t)^{A-n}\,.\label{e:pn}
\end{equation}

Here you have the expression $\sigmain t$ raised to the
$n$\th~power, that is the probability  for $n$ particular
collisions to take place with $n$ nucleons. Then for the remaining
$A-n$ nucleons such collisions must not happen. Eq.~(\ref{e:pn})
simply counts all the ways these events can take place.

Now I want to generalize this model in a way that may interest
you. I am going to assume that there is a number $\mu$  which is a
kind of efficiency in a single collision for doing whatever
process interests us. So let us say, the first collision, on a
virgin nucleon produces one unit of whatever we are looking for.
The second one only produces a fraction of that, a fraction we
shall call $\mu$. The third one produces $\mu^2$, and so on. My
Polish friends, when $\mu = 0$, called this the wounded nucleon
model~\cite{6}. I do not know what to call this more general
model, but I think that by analogy with the way ``pre-owned" cars
are traded in America, this might be called the ``used nucleon
model." If you ask what is the number of particles that get
produced in these $n$ collisions, well, you have a finite
geometrical series to sum, and here is the answer:
\begin{equation}
1+\mu+\dots+\mu^{n-1} = \frac{1-\mu^n}{1-\mu}\,.
\end{equation}

Now let us put the pieces together. It is convenient to abbreviate
the cross-section times the thickness function, by calling  it $x
= \sigmain t(b)$. The average multiplicity of the event weights
that finite geometrical sum with the probabilities we derived a
moment ago. Now the steps that follow are just uninteresting
algebra, but the result is very simple. What it does is to
multiply the individual cross-sections by the number $1-\mu$, and
then it renormalizes the entire expression by a factor of
$1/\left(1 - \mu\right)$.
\begin{eqnarray}
\textnormal{Av. multiplicity }&=&\Sigma_nP_n\frac{1-\mu^n}{1-\mu}\\
&=&\frac{1}{1-\mu}\Sigma\combin{A}{n}x^n(1-x)^{A-n}(1-\mu^n)\nonumber\\
&=&\frac{1}{1-\mu}\left\{(x+1-x)^A-(\mu x+1-x)^A\right\}\nonumber\\
&=&\frac{1}{1-\mu}\left\{1-\left[1-(1-\mu)x\right]^A\right\}.\nonumber
\end{eqnarray}

There are two limits for this expression.  The one extreme, $\mu =
0$, is pretty dramatic. It corresponds to a kind of automobile we
once had  in America called the Yugo. One collision and it is
gone. In that case, the average multiplicity is just the same as
the probability of at least one collision. And what you have there
is the wounded nucleon model of my friends Bialas, Bleszinski, and
Czyz~\cite{6}. On the other hand, if you take $\mu = 1$ or, more
correctly, let $\mu$ approach one as a limiting process, then the
average multiplicity turns out to be $Ax$. It is just the average
number of collisions in which you count all of them with equal
weighting.

Now what does this mean when you have two nuclei colliding with
one another? It yields a somewhat longer formula, but that  is
just the formula you have seen before for the wounded nucleon
model except that it has some factors of $1-\mu$ inserted:
\begin{eqnarray}
\av{\textrm{Mult}}_\mu &=&
\frac{A}{1-\mu} \int T_A(s) \left\{1-\left[1-(1-\mu)\sigmain
T_B(b-s)\right]^B\right\}d^2s\nonumber\\
&+&\frac{B}{1-\mu} \int T_B(s) \left\{1-\left[1-(1-\mu)\sigmain
T_A(b-s)\right]^A\right\}d^2s.
\end{eqnarray}

So the calculation is always very much the same as the one that
has been done by each of the RHIC groups. However, there are the
renormalizations by factors of $1 - \mu$ involved, and in
particular for $\mu = 0$ the expression before us is just what has
been called the number of participants, or in Poland the number of
wounded nucleons. On the other hand when we let $\mu$ go to one,
what it becomes is twice the total number of collisions. Not the
number of collisions, but twice it, because every collision has
two participants. It is the number of collided nucleons rather
than the number of collisions. So this ``used nucleon" model gives
you a way of interpolating between the two limits that are
discussed most, and it associates perhaps a new parameter $\mu$
with whatever sort of process interests you.

Now I want to change the subject altogether and begin to talk
about quantum optics. For that, let me remind you of a famous sort
of astronomical experiment first performed by two men trained as
radio engineers, R. Hanbury Brown and R. Q. Twiss~\cite{7}. They
used two radio antennas pointed at a radio source and chose to
detect each signal separately to take away its high frequency
component. They felt it would be much easier then to handle the
remaining low frequency modulations that are present in such an
astronomical signal. They sent those signals to a central point
where they were multiplied together and averaged over time.

How does one analyze an experiment of that sort? It is very easy
to show that there is an interference term present in the product
of these two detected signals. That interference term is rather
different from the interference terms one ordinarily encounters in
19\th century optics.  Let us divide the field into two complex
conjugate pieces, and talk then about the two complex fields that
each oscillate with a single sign of the frequency:
\begin{equation}
E(r,t) = E^{(+)}(r,t)+E^{(-)}(r,t)\,\textrm{, and}
\end{equation}
\begin{equation}
E^{(-)}(r,t)=\left\{E^{(+)}(r,t)\right\}^*\,.
\end{equation}

Virtually all of traditional wave optics, that is essentially all
of optics until the early 1950's then can be discussed in terms of
the lowest order field correlation function, which is an averaged
product only quadratic in the field strengths:
\begin{equation}
G^{(1)} \equiv \av{E^{(-)}(r,t)E^{(+)}(r',t')}\,.\label{e:g1}
\end{equation}

But Hanbury Brown and Twiss were telling us that you have to deal
with a quartic expression, because you have two detected signals
each of which is quadratic in the field strengths:
\begin{equation}
G^{(2)} \equiv
\av{E^{(-)}(r,t)E^{(-)}(r',t')E^{(+)}(r',t')E^{(+)}(r,t)}\,.\label{e:g2}
\end{equation}

That led them, I'm afraid, to a terrible dilemma. They were eager
to perform the same interferometry  with visible  light as they
had with the radiofrequency fields that they regarded as
classical, but for a time fell victims to a great confusion about
the quantum theory which has not entirely disappeared. To do the
same sort of experiment with light quanta, they realized you
obviously need to annihilate two quanta, one at each detector.
When you read the first chapter of Dirac's famous textbook in
quantum mechanics~\cite{8}, however, you are confronted with a
very clear statement that rings in everyone's memory. Dirac is
talking about the intensity fringes in the Michelson
interferometer, and he says,

\begin{quote}
Every photon then interferes only with itself. Interference
between two different photons never occurs.
\end{quote}

\begin{figure}
\begin{center}
  \includegraphics[width=220pt]{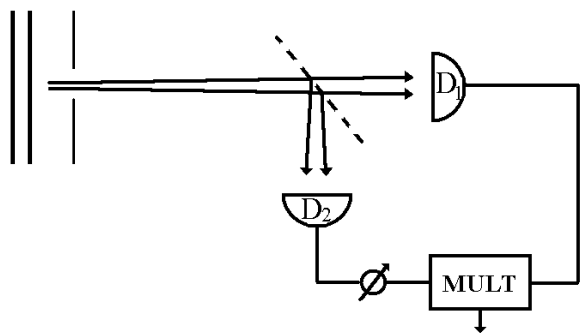}\\
  \caption{
Setup of the Hanbury Brown -- Twiss experiment. Light from an
extremely monochromatic discharge tube goes to a half-silvered
mirror, which sends beams to two separate photo-detectors. The
random signals coming from those two detectors are multiplied
together and then averaged.}\label{f:hbt}
\end{center}
\end{figure}

Now that simple statement, which has been treated as scripture, is
absolute nonsense. First of all, the things that interfere are not
the photons themselves, they are the probability amplitudes
associated with different possible histories. You can obviously
have different histories that involve more than one photon at a
time.   The intensity interferometry of Hanbury Brown and Twiss
was indeed exploiting the interference of pairs of photons. But
they were not completely persuaded, and I am not sure they were
ever fully persuaded. Instead they chose to resolve the dilemma by
doing another experiment that was both cleverer and more
revolutionary than the experiments on determining the angular
sizes of radiofrequency sources. Here, in Fig.~\ref{f:hbt} is
their experimental setup~\cite{9}. Their light source is an
extremely monochromatic discharge tube. The light from that goes
to a half-silvered mirror, which sends beams to two separate
photo-detectors. The random signals coming from those two
detectors are multiplied together, as they were in their
radiofrequency experiments and then averaged. What they found was
a slight tendency for both of these photodetectors to go off
together as they varied the position of one of them and thus the
time delay between them. It was only a slight tendency toward
correlation because in fact the effect was spread out over time by
the rather poor resolution time of their detection equipment. The
experiment was later repeated somewhat more accurately by Pound
and Rebka~\cite{10}. Subsequent repetitions of this experiment
have indicated that if you have very high resolving power in your
counters, you can detect something approaching an increase of a
factor of two in the coincidence rates for very short time delays.

Some nonsensical things were initially said about the
interpretation of this experiment, but a correct interpretation
was soon given by Purcell\cite{11} on the basis of semi-classical
reasoning and a classical formula for noise power he had quoted
from the Radiation Laboratory Handbook.  But there were other
papers published that still misinterpreted the effect. One said
that because of the very narrow linewidth and coherence
characteristic of a laser beam, there should be a very dramatic
Hanbury Brown -- Twiss correlation in such a beam. I was very
puzzled by that claim and published a brief note saying no, there
will be absolutely no such effect at all in a laser beam, in fact
no photon correlations whatever. But then that led back to the
question, what is coherence? Optical coherence as it is usually
construed deals only with the first or lowest order correlation
function of Eq.~(\ref{e:g1}), which is quadratic in the fields.
You can define higher orders of coherence, in which the higher
order correlation functions such as given by Eq.~(\ref{e:g2}) will
factorize. When they do factorize it wipes out the effect of
photon bunching. The Hanbury Brown -- Twiss correlation effect
disappears completely. Well, all of the $G^{(n)}$'s can factorize,
and in that case you can speak of full coherence. That holds in
particular for what I have called the coherent states because they
are eigenstates of the photon annihilation operators. When you
annihilate a photon in those states  the state doesn't change. The
annihilation operator just multiplies the state by a constant so
that the correlation functions of all orders just factorize.

I had been able to show back in the early 50's~\cite{13}, that any
classical or pre-determined current will radiate coherent states.
That raises the question then, what is the effective current for a
laser? The answer is, that in a laser there is a strongly
oscillating transverse polarization current that radiates its
beam. It is a very powerful current, but the electric charge does
not go anywhere. It remains bound in the gas atoms. That in a
nutshell is why a laser beam does not have a Hanbury Brown --
Twiss correlation for its photons.

There is a fair-sized field that developed since then, which we
now call quantum optics. In a sense it is really the study of
time-dependent photon statistics. I would like to show you now in
a minute or two that photon statistics can have something to say
about heavy ion collisions. There is a very close parallel in some
ways between the statistical analysis you do for emitted bosons,
and the problems of quantum optics.

Let us suppose we are presented with a pure signal, which is
represented by a coherent state $|\beta\rangle$. If you try
counting the number of photons in that state for any length of
time, you will always obtain a Poisson-distribution about the mean
value $|\beta|^2$\,:
\begin{equation}
p(n) = \frac{|\beta|^{2n}}{n!}\textrm{e}^{-|\beta|^2}\,.
\end{equation}

Suppose, on the other hand, you are presented with a signal that
consists of pure noise. That is to say you have a Gaussian
distribution
\begin{equation}
P(\alpha) =
\frac{1}{\pi\av{n}}\textrm{e}^{-\frac{|\alpha|^2}{\av{n}}}
\end{equation}
of the amplitudes of coherent states. For that case you obtain a
famous geometrical distribution for the number of photons that you
count:

\begin{equation}
p(n) = \frac{1}{1+\av{n}}\left(\frac{\av{n}}{1+\av{n}}\right)^n\,.
\end{equation}

What happens when you superpose noise and signal? Well in the
coherent state language it is very simple. You just displace the
Gaussian distribution function by $\beta$. But when you do the
calculation of the $n$-quantum probabilities, the distribution of
the number of quanta you find is no longer so simple:
\begin{equation}
p(n) = \left(\frac{\av{n}}{1+\av{n}}\right)^n
L_n\left(-\frac{|\beta|^2}{\av{n}(1+\av{n})}\right)
\textrm{e}^{-\frac{|\beta|^2}{1+\av{n}}}\,.
\end{equation}

\begin{figure}
\begin{center}
  \includegraphics[width=220pt]{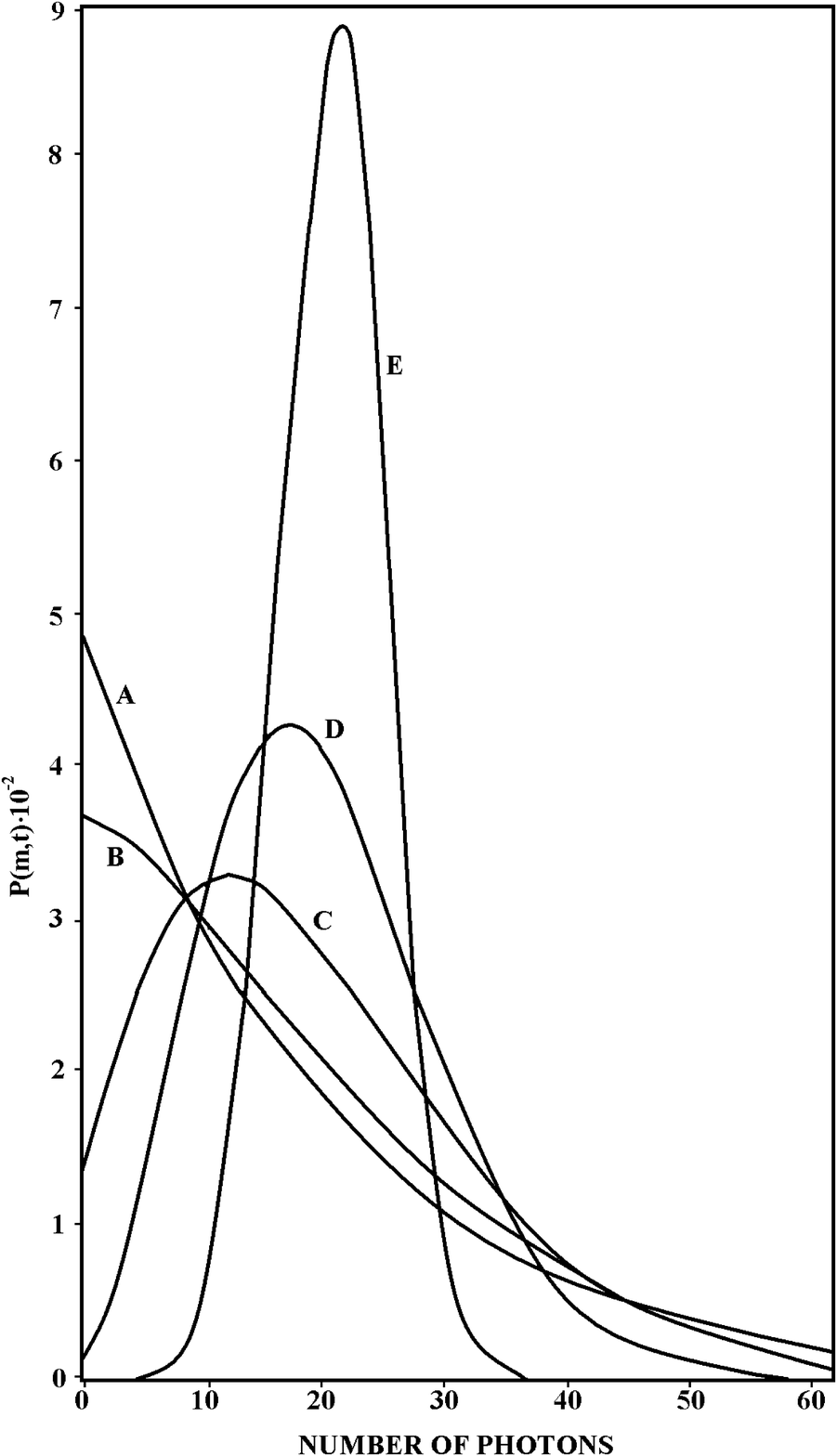}\\
  \caption{Photon count distributions which would be measured in various
superpositions of coherent (signal) and chaotic (noise) fields.
Curve A represents the distribution for a pure noise field. The
other curves represent the distribution for fields in which the
noise and signal components would separately contribute the
following average numbers: Curve B, 10 from noise and 10 from the
signal; Curve C, 5 and 15; Curve D, 2 and 18; Curve E, 0 and 20,
respectively.}\label{f:numdist}
\end{center}
\end{figure}

There is nothing obvious about this result at all. The $L_n$ in it
happens to be the $n$\th~Laguerre-polynomial.
Figure~\ref{f:numdist} shows you what that distribution looks like
if the average number of quanta counted is 20. If the signal is
purely coherent, you have the familiar Poisson distribution E. If
it is pure noise on the other hand, you have the geometrical
distribution, given by curve A. Curve B corresponds to half signal
and half noise, and it still looks a great deal like the noise
distribution. Curve D corresponds to corrupting the signal with
only two quanta on the average. You have an average of 18 quanta
from the coherent signal, and just two quanta of noise.  A little
noise goes a long way toward changing  the distribution.

Let me now ask, can there be any coherence present in the
heavy-ion output of pions, for example?  Aren't the violent
collisions you study just about the most incoherent things you
could possibly imagine? Well, let us consider the example of the
simplest of all meson theories. We take the interaction to be that
between a meson field $\phi(r,t)$ and a fixed source $\rho(r)$:
\begin{equation}
H_\textrm{int} = \int \rho(r) \phi(r,t) d^3r\,.
\end{equation}

The ground state in that theory is in fact a coherent state,
$\Pi_k|\alpha_k\rangle$.  It happens to be a bound one; the pions
can't go anywhere. The amplitudes are as follows:
\begin{equation}
\alpha_k\propto\frac{\int\rho\textrm{e}^{-ikr}d^3r}{E(k)}\,,
\end{equation}
where $E(k)$ is the energy for the $k$\th~mode. Now if you were
suddenly to wipe out this source density, that is to set it
instantly equal to zero, what would happen? Of course you would
have to supply energy to the field to do that. The coherent
excitation would suddenly be set free, and you would have a
coherent state of free pions, with no momentum correlation at all
between them. Now suppose we don't do anything so drastic as to
wipe out the source $\rho$ but instead we replace $\rho$ suddenly
by a wildly random source instead of a smooth source of any sort.
That random source is going to produce random mode excitations,
and the density operator that describes the field is going to be
some sort of a distribution over a set of random amplitudes
$\gamma_k$. The random excitations superposed upon the initially
coherent state will lead to a density operator

\begin{equation}
\int P(\left\{\gamma_k\right\})\Pi_k
|\alpha_k+\gamma_k\rangle\langle\alpha_k+\gamma_k|d^2\gamma_k.
\end{equation}

Now the coherent excitation remains, and you can't get rid of it.
It remains there, but its contribution could be altogether swamped
by the vast amount of noise an energetic collision adds to the
field. The answer is nonetheless: Yes, you can indeed have some
coherence present in this extremely incoherent output.  It could
be one of several reasons why the measured correlation functions
for like pions do not rise to the value 2 for vanishing relative
pion momenta.

\vfill\eject

\end{document}